# Thermalization and long-lived quantum memory in the multi-atomic ensembles

S. A. Moiseev[1,2*], and V. A. Skrebnev[2**]

[1] Quantum Center, Kazan National Research Technical University,

10 K. Marx, Kazan, 420111, Russia

[2] Zavoisky Physical-Technical Institute of the Russian Academy of Sciences,

10/7 Sibirsky Trakt, Kazan, 420029, Russia

E-mails: samoi@yandex.ru*, vskrebnev@mail.ru**

We have studied the role of thermalization in lifetime reduction of quantum memory based on multi-atomic ensembles. Herein, it is shown to be impossible to remove the thermaization-caused decoherence in such systems using the methods of dynamical decoupling. We have analyzed the existing models of the thermalization and have proposed a new understanding of the thermalization as a result of the internal processes which are not described in the unitary quantum dynamics formalism. The possible ways for reducing the negative influence of thermalization on the quantum memory lifetime are also discussed.

**Keywords.** Quantum memory, decoherence, dynamical decoupling, thermalization, canonical distribution.

## 1.Introduction

Creation of quantum memory (QM) is a topical problem in realization of long-distance quantum communication and creation of universal quantum computer [1-3]. Now great expectations are connected with using macroscopic atomic ensembles [4-8] for creation of quantum memory devices. The photon echo QM approach [7,9-19] provides an efficient quantum storage of the multi-qubit light fields. This approach is based on the controllable realization of perfect time-reversible dynamics of multi-atomic systems [7,9,16,19]. The record results [11,13-15,18] and promising properties for quantum storage of many photonic qubits have been recently demonstrated for solid state media [10,12]. Nitrogen-vacancy centers in diamond [20-24] and phosphorus ions in silicon [25,26] are also considered for application in quantum storage.

In spite of large progress, the developed QM schemes require deeper studies of the fundamental problems in quantum dynamics of multi-atomic systems since the quantum storage devices require nearly perfect time-reverse dynamics of the macroscopic systems and strong suppression of all the decoherence effects in these systems. Here we discuss some of these problems related to the atomic ensembles characterized by significant interactions of atoms. Implementation of perfect time-reversible dynamics for the atomic systems assumes studying the irreversibly sources in the macroscopic systems. Absence of sufficiently complete understanding of these problems determines the basic limits in the implementation of highly efficient QM-protocols and longer QM lifetime.

Realistic possibility for the QM lifetime increase is associated with transfer of the mapped flying photonic qubits to the long-lived electron and nuclei spin states [27,28]. The best results are achieved by using the experimental methods leading to active dynamic suppression of the decoherence effects (the so-called dynamical decoupling (DD)-technique) in the nuclei and electron spin systems [29-31]. Success of this approach is explained by realization of controllable reversible unitary dynamics in the complicated quantum systems.

Clearly, the multi-particle systems are the most difficult and important objects requiring detailed theoretical and experimental investigations.

*1.1. Multi-pulse sequences for the dynamical decoupling*

Originally, the DD methods were proposed during the development of the high-resolution multi-pulse nuclear magnetic resonance (NMR) methods [32,33]. Then, this approach was applied in the pulsed NQR-experiments on the system with the quadrupole moment of the nuclear spins [34] and has been successfully demonstrated on the system of electron spins [35]. Recently new DD pulse sequences were proposed for the suppression of the decoherence effect in the qubit evolution caused by the fast fluctuating spin-bath and pulse imperfections [36]. The well-known Carr-Purcell (CP) radio-frequency (rf-) pulse sequence and its modification - Carr-Purcell-Meiboom-Gill (CPMG) sequence demonstrate very effective suppression of the local field inhomogeneities [37,38]. Also, it is worth noting XY- sequence [39] which is an improved CP - pulse sequence based on "x" and "y" phase alternation of the rf-pulses. The sequence compensates cumulative pulse errors for all three components of magnetization. In the recent work [40], the authors have proposed the modified CPMG sequence with chirped rf-pulses. The sequence operates with greater bandwidths and provides for significant signal-to-noise improvement when compared to the standard CPMG sequence.

To counter the effect of the environment on the quantum system, the continuous DD could be also effective, which uses continuous driving rf-fields. It has been demonstrated in [41] that the continuous driving fields can significantly decouple spins from magnetic noise. In work [42], a field configuration utilizing local static fields and few continuous driving fields are constructed for achieving the protection against decoherence of the two-qubit states.

Waugh- Huber- Haeberlen (WAHUHA) and Mansfield-Rhim-Elleman-Vaughan (MREV) sequences [43,44] are widely used for suppression of the decoherence caused by the dipole-dipole interactions in the multi-atomic ensembles.

The composite variants of the above sequences with 34 pulse cycle are used for decoherence reduction in the system of nuclear spins due to the simultaneous presence of the dipole-dipole interactions and the local field inhomogeneities [23,45]. However it is desirable to have shorter DD pulse sequences leading to suppression of the main decoherence sources, since a large number of pulses in the sequences limits their efficiency in practical implementations.

Recently we have proposed new pulse sequences for suppression of the decoherence effects caused by the simultaneous presence of the dipole-dipole interactions and the inhomogeneities of magnetic fields in the spin systems [46,47]. It has been shown that the specific choice of the pulse sequences allows using a significantly smaller number of pulses in the sequences compared to the number of pulses in the composite sequence used in [23,45]. The proposed technique [46,47] demonstrates a new possibility for the efficient increase in the lifetime for multi-qubit QM in the concentrated multi-atomic ensembles.

*1.2. Dynamical decoupling and irreversibility*

The action of all DD pulse sequences is based on the minimization of the Hamiltonian terms responsible for the decoherence of the system which develops according to the von Neumann equation

$$\dot{\rho}(t) = -[H(t), \rho(t)]. \qquad (1)$$

This minimization is achieved by reversing the sign of the corresponding members of the Hamiltonian at different stages of the system evolution. For example, when CP sequence [37] is used, the sign of the interaction of the spin systems with inhomogeneous local fields is reversed. If WAHUHA and MREV sequences [43,44] are used, the sign is reversed for the operators describing dipole-dipole interactions in the system.

As the general solution of the Schrödinger equation

$$i\hbar \frac{\partial \psi}{\partial t} = \hat{H}\psi, \tag{2}$$

has the following form

$$\psi(t) = e^{-i\frac{\hat{H}}{\hbar}t}\psi(0), \tag{3}$$

it is evident that the Hamiltonian sign reversal is identical to time sign reversal. Hence, sign reversal of the entire Hamiltonian would be supposed to return the system to its initial state and decoherence would be supposed to disappear. However, in the macroscopic systems of the interacting atoms we will face the problem of irreversibility. In turn, the irreversibility is linked to thermalization, which allows us to describe such systems by the methods of statistical physics. Here we need to note that there is no foundation for a belief that irreversibility and thermalization are connected to the so-called dynamical chaos [48,49], since the mechanism of dynamical chaos is absent in quantum mechanics due to the strictly linear time evolution of the Schrödinger equation.

At present, the irreversible behavior of macroscopic systems remains unexplained in terms of unitary quantum dynamics and there are no reasons to claim that Hamiltonian sign reversal will completely reverse the evolution of macroscopic systems. The experiments [50,51] have used the pulse sequences which changed the sign of Hamiltonian with predetermined accuracy in the macroscopic system of dipole-coupled nuclear spins situated in the external magnetic fields. This has allowed for the irreversible components of the system evolution to be singled out and investigated.

Thus, there may exist irreversible processes in the macroscopic systems which are not reflected in the formalism of the unitary quantum dynamics described by the Schrödinger equation. These processes can determine the irreversibility of the evolution and thermailzation of the macroscopic systems. They will be the main cause of decoherence if the perfect DD methods have removed all other causes. So, the long-time preservation of non-equilibrium quantum coherence in the multi-atomic ensembles (storing quantum information in the multi-atomic ensembles) faces the basic physical problems of the multi-particle dynamics and thermalization. This poses new challenges in the solution of these basic problems.

If we understand the true causes of thermalization, we will be able to minimize the decoherence effects caused by the thermalization. In the next section, we discuss the problems of existing models of thermalization. Then we describe the thermalization as the result of internal processes characteristic for the macroscopic systems, and we discuss possible reduction of the negative effects of thermalization on the QM lifetime.

## 2. The models of thermalization

1. It is well-known that thermalization of the arbitrary initial state of macrosystem leads to the state described by Boltzmann-Gibbs distribution (or canonical distribution):

$$\rho(E_n) = \frac{e^{-\beta E_n}}{\sum_n e^{-\beta E_n}}. \qquad (4)$$

Two models are used for the derivation of the canonical distribution (e.g. see [52]). In one of them, the system under consideration is assumed to be a sub-system of a very large system; the environment of the sub-system is often called the thermostat. But since the boundaries of the thermostat are unknown, as it is unknown what is beyond them, the total system is called the Universe, modestly placed in quotation marks – "Universe". The "Universe" is postulated to be in equilibrium.

In another model, the "Universe" is assumed to consist of an enormous number of systems identical to the system under consideration. The most probable distribution of those systems on energy levels turns out to correspond to canonical distribution in the system under consideration.

Both models use only the system's eigenstate as the state of the system with energy E. Both assume that the system's interaction with environment is negligibly small, which means the system is considered practically isolated. However, the solution of the Schrödinger equation (2) for such system can be written as:

$$\psi = \sum_n c_n(t)\psi_n, \qquad (5)$$

where $\psi_n$ are eigenfunctions of system Hamiltonian,

$$c_n(t) = c_n(0)\exp(-\frac{i}{\hbar}E_n t). \qquad (6)$$

In accordance with (5), the total system energy equals

$$E = \sum_n |c_n|^2 E_n. \qquad (7)$$

Normalization requirement gives us the following:

$$\sum_n |c_n|^2 = 1. \qquad (8)$$

If the number of the energy levels is more than two, equations (7) and (8) have a great number of solutions for $|c_n|^2$. This means that the same value of energy $E$ corresponds to the great number of different system states.

As neither the Universe, nor "Universe" are in equilibrium, nor do they consist of the great number of systems identical to the system under consideration, nor only the eigenstates of the system are possible, we can say with good reason that both above models used for the derivation of canonical distribution are artificial and have no relation to the physical reality. Accepting an incorrect derivation of a correct formula prevents the true understanding of the processes which the formula describes.

2. Articles [53-55] look at the ensemble of system states with the same energy. These states are considered as equiprobable. The entire set of equiprobable states is called the *Generalized Quantum Microcanoncal Ensemble* (GQME).

Since in quantum mechanics the probability of the energy being equal to $E_n$ is determined by the value $|c_n|^2$, one might have supposed that in a macroscopic system the value of $|c_n|^2$ averaged by GQME would correspond to canonical distribution. In article [55] all quantum superpositions of form (5) satisfying condition (7) are considered to be equally probable. However, the averaging by this manifold performed in [55] has yielded the values of $(|c_n|^2)_{av}$ in a significant departure from the Boltzmann-Gibbs statistics. This indicates that using GQMA does not yield results corresponding to the thermalization of the system.

The authors of works [53-55] do not give a physical substantiation for the possibility of averaging by GQME. At the same time the new approach to description of macrosistems proposed in [53-55] has demonstrated existing problems of standard models of statistical mechanics.

3. Articles [56-58] study the thermalization of an isolated system on the basis of the so-called *Eigenstate Thermalization Hypothesis* (ETH). The ETH is a set of ideas which purports to explain when and why an isolated system can be accurately described using equilibrium statistical mechanics. In particular, it is devoted to understanding how systems which are initially prepared in far-from-equilibrium states can evolve in time to the state of thermal equilibrium.

The ETH contains a number of assumptions, which allow its authors to approach the desired result. It is assumed that the initial state of the system (which is far-from-equilibrium) is some superposition of energy eigenstates which are all sufficiently close in energy, that is are located in a narrow energy window. The ETH says that for an arbitrary initial state, the expectation value of some quantum-mechanical observable $\hat{A}$ will ultimately evolve in time to its value predicted by a microcanonical ensemble. Thereafter $\hat{A}$ will exhibit only small fluctuations around that value, provided that the following two conditions are met: 1) the diagonal matrix elements $A_{nn}$ vary smoothly as a function of energy, with the difference between neighboring values, $A_{n+1,n+1} - A_{nn}$ becoming exponentially small in the system size; 2) the off-diagonal matrix elements $A_{nm}$ are much smaller than the diagonal matrix elements, and in particular are themselves exponentially small in the system size.

The explicit expectation value of any observable $\hat{A}$ at any given time is

$$\bar{A} = \langle\psi(t)|\hat{A}|\psi(t)\rangle = \sum_{n,m} c_n^* c_m A_{nm} e^{-i(E_m-E_n)\frac{t}{\hbar}}. \tag{9}$$

However, instead of (9) the ETH assumes a long-time average of the expectation value of the $\hat{A}$:

$$\bar{A} = \lim_{\tau\to\infty}\frac{1}{\tau}\int_0^\tau \langle\psi(t)|\hat{A}|\psi(t)\rangle\, dt = \lim_{\tau\to\infty}\frac{1}{\tau}\int_0^\tau \sum_{n,m} c_n^* c_m A_{nm} e^{-i(E_m-E_n)\frac{t}{\hbar}}\, dt$$

$$= \sum_n |c_n|^2 A_{nn}. \tag{10}$$

The assumptions of the ETH not have a proper physical explanation. Currently, there is no known analytical derivation of the ETH. Thus, the ETH existence also indicates that the problem of thermalization of macrosystems should be solved.

## 3. Thermalization as a result of internal processes in macroscopic systems

Here we offer a new derivation of canonical distribution that is free from assumptions obviously contradicting to the physical reality. The proposed derivation is based on the Boltzmann's method of the most probable distribution and takes into account the internal processes which exist in physical systems and result in canonical distribution and, consequently, in thermalization.

Remember that according to quantum mechanics the value $|c_n(t)|^2$ in the equation (7) determines the probability of system energy being equal to $E_n$. In accordance with (6) we have:

$$|c_n(t)|^2 = |c_n(0)|^2. \tag{11}$$

We see that quantum mechanics does not allow the system to pass into a state with a set of $|c_n(t)|^2$ different from the initial set. However, experience shows that in a macrosystem left to its own devices after an impact inducing certain initial conditions, the probability of the system's having energy $E_n$ after some time (the relaxation time) becomes described by canonical distribution, i.e. it does change. The speed of arriving at the canonical distribution does not depend on the properties of the surface of the macrosystem, nor on the structure of its environment. Thus, the influence of environment does not explain the transition of the probabilities of the system being in a state with energy $E_n$ to canonical distribution. This means that there must be internal processes which determine the transition of the initial distribution of probabilities to the canonical distribution. Hence, the canonical distribution may be derived as a result of internal processes of the macrosystem.

The solution of Schrödinger equation is called the wave function. This function determines the distribution of probability of the values of the system's physical characteristics (e.g. the particle coordinates or energy). Quantum mechanics says nothing about internal processes within the system which provide for the probability distribution. Therefore, we shall call them hidden processes. It is obvious that if these hidden processes did not exist, probability distribution would not exist either.

The possibility of finding a particle in a certain point of space corresponds to one of the "instantaneous" states caused by the internal processes in the system. The wave function, which is a solution of the Schrödinger equation, allows to find the probability of those states. Since we are talking about probability of certain states, there must be transitions between those "instantaneous" states. Because transitions between "instantaneous" states exist, the states themselves also must exist.

Hidden internal processes in physical systems are very fast (recently [59,60] have shown that the lower boundary of the speed of the Einstein's "spooky action at a distance" is $10^4$ light speeds). It is clear that a wave function may be used to describe the states of physical systems because the hidden processes are extremely fast. This means that to describe the hidden internal processes we need a time scale whose graduation is many orders smaller than the usual.

It follows from the above that quantum mechanics formalism describes the averaged state of the quantum system over times that are considerably longer than the times taken by the hidden internal processes. This is why the "instantaneous" system states and the mechanism of moving between those states end up beyond the quantum mechanical description.

According to quantum mechanics the value $|c_n(t)|^2$ in the equation (7) determines the probability of system energy being equal to $E_n$. Obviously, some physical state of the system must correspond to every different $E_n$. It is only possible to treat $|c_n|^2$ as a probability of the system being in a state with energy $E_n$ if the system transfers between the states with different $E_n$. Clearly, formula (7) and the wave function (5), containing $|c_n(t)|^2$ and $c_n(t)$, may be used only under extremely high frequency of transitions between the states with different $E_n$. Consequently, every state with energy $E_n$ appears in the system for a very short time. We will use the term "$E_{ni}$-states" for such "instantaneous" states with energy $E_n$. Importantly, $E_{ni}$-states should not be confused with the system's eigenstates which are described by wave functions – the eigenfunctions of the system's Hamiltonian.

Transitions between $E_{ni}$-states, just as Einstein's spooky actions, are not described by quantum mechanics, but it does not mean that they do not exist in the physical reality. We call this type of hidden internal processes *ghost actions*. Because transitions between $E_{ni}$-states exist, the states themselves must exist in physical reality. $E_{ni}$-states which correspond to the same energy value may differ in other parameters. The values of $|c_n|^2$ are proportional to the average time of the system being in the $E_{ni}$ – states.

To derive canonical distribution, we will use the method of the most probable distribution [52,61], but instead of the Universe consisting of an enormous number of macrosystems identical to the system under consideration we will consider a great number of events which take place in the single macroscopic system under consideration. Each event is a visit of the system to one of the "instantaneous" states with energy $E_n$.

Let $N$ be the number of the system's cumulative visits of its "instantaneous" energy states over time $t$ and let $\nu_n$ be the number of visits of $E_{ni}$-states, corresponding to the energy $E_n$, over this time. Obviously,

$$N = \sum_n \nu_n. \tag{12}$$

Let's introduce the value

$$E_t = \sum_n \nu_n E_n. \tag{13}$$

Numerous "configurations" determined by various sets of numbers of visits $\nu_n$ correspond to the value $E_t$. Each configuration may be realized in $P$ ways corresponding to the number of permutations of the visits:

$$P = \frac{N!}{\nu_1! \nu_2! \ldots \nu_l! \ldots}. \tag{14}$$

Assuming the number $N$ to be very large, we find the maximum of the function $P$ under conditions (12) and (13) and arrive at the most probable value of the numbers of visits of the $E_{ni}$-states:

$$\nu_n = \frac{N e^{-\beta E_n}}{\sum_n e^{-\beta E_n}}. \tag{15}$$

The probability $\rho(E_n)$ of the system being in the $E_{ni}$-states equals the ratio of the number of the visits of those states to the total number of visits $N$:

$$\rho(E_n) = \frac{\nu_n}{N} = \frac{e^{-\beta E_n}}{\sum_n e^{-\beta E_n}}. \tag{16}$$

Thus, we have arrived at the canonical distribution. The value $\beta$ is determined by the equation

$$E = \sum_n \rho(E_n)E_n. \tag{17}$$

The irreversible process of the initial state of the macrosystem turning into a state with the most probable distribution (15) of the values $\nu_n$ and canonical distribution $\rho(E_n)$ (16) is only possible if there exist transitions between $E_{ni}$-states which are not described by quantum mechanics and which provide for this change while the total energy $E$ of the system is preserved. Let's call those transitions "relaxation ghost action", or "relaxation transitions" (r-transitions).

The system state where the distribution of probability is determined by the Schrödinger equation is called pure state. If the distribution of probability is determined by formula (16), which corresponds to the thermalization of macrosystem, such state is called mixed state (see, e.g. [62]). To describe the mixed state we must use the density matrix, in which the non-diagonal components are assumed to equal zero in the energy representation, and diagonal components are described by (16). It is natural to consider that in the closed macrosystems there exist the processes breaking the unitary Schrödinger evolution and leading to the mixed state of the macrosystems.

Appearance of mixed states in macrosystems may be viewed as some transformation or reduction of the macroscopic system wave function. The irreversible process of this transformation (reduction) itself is not described by the Schrödinger equation.

It is known that the interaction of a particle with the macroscopic system under certain conditions leads to the reduction of the particle's wave function. As the result of the reduction, the particle (e.g. a photon interacting with the macroscopic screen) can be found in a certain place with the probability determined by its wave function. The process of such reduction also is not described by the Schrodinger equation.

Thus, the Schrödinger equation cannot correctly describe the irreversible processes taking place in macrosystems. It is obvious that a macrosystem is a system because its particles interact with each other. If the number of objects (e.g. particles) in a system is small, Schrödinger equation describes the system correctly and the irreversibility of the system's evolution does not manifest noticeably. But when the number of particles is macroscopic, the number of their interactions with each other is huge and irreversibility becomes a fact. At present it is unknown which interactions and why lead to the appearance of mixed states in macroscopic systems or to the collapse of the wave function of one particle when it interacts with a macro object. To answer these questions, a deep investigation of limitations of quantum theory in describing macroscopic systems is called for.

This paper points out that scientists should take into account those processes in the physical systems which are not described by the unitary Schrödinger evolution. We have shown that the fundamental formula of canonical distribution can be received through the method of most probable distribution which accounts for such processes as relaxation ghost action, and does not require any physically unjustified assumptions.

Using equations (14) and (15) it is easy to show that for maximum $P$

$$lnP_{max} = -N \sum_n \rho(E_n) ln\rho(E_n) = N(\beta E + ln \sum_m e^{-\beta E_m}) = \frac{N}{kT}(E - A), \quad (18)$$

where $A$ is the free energy of the system.

Whereas $E - A = TS$, the entropy of the system is equal to

$$S = k\frac{lnP_{max}}{N} = -k \sum_n \rho(E_n) ln\rho(E_n) = -kln\rho(E). \quad (19)$$

Entropy is one of the fundamental concepts of science. Formula (19) connects entropy to the maximum number of system state realizations as a consequence of hidden internal processes which provide for thermalization in physical systems. The macrosystem may be said to tend towards maximum freedom in realizing its states with a given total energy, i.e. towards maximum entropy, expressed by (19).

## 4. Conclusion

We think that after decoherence processes are suppressed in macrosystems as much as possible by the dynamical decoupling methods, thermalization will become the main source of the remaining decoherence. According to our concept, thermalization is the result of extremely fast processes which take place in macrosystems and are not reflected in the existing quantum mechanics formalism.

Probing the foundations of quantum mechanics has been a long-standing goal since the pioneering work of John Bell on nonlocality. Quantum memories (based on single trapped atoms, or ions) offer a promising approach towards this goal [8].The study of the influence of the irreversible component of macrosystem evolution on quantum memory will also yield valuable information about the processes which are not described by unitary mechanics and which provide for the thermalization and decoherence process in the multi-atomic systems.

If a system follows the laws of thermodynamics, its total energy determines its temperature. In this case, the type of objects forming the thermodynamic system depends on its temperature. At certain temperatures a phase transition occurs from one type of system-forming objects to another. For example, in the paramagnetic phase the system is formed by the magnetic moments of atoms. When the temperature drops, the magnet transitions into the ferromagnetic phase, where a system of magnons is formed, which obeys the laws of thermodynamics. When the energy (i.e. temperature) rises, the gas containing atoms and molecules turns into plasma and so on. During thermalization, a system transitions to a state of thermodynamic equilibrium while maintaining its energy.

A group of objects (for instance, of nuclear spins) is a system because they interact with each other. It is natural to assume that thermalization of the system is connected to the system Hamiltonian terms responsible for the interactions. The unanswered question is what is the nature of this connection? The answer will allow us to look for the macrosystems with the maximal thermalization time and, accordingly, maximum decoherence time. An empirical search may also be productive.

When the number of particles goes down, the processes causing thermalization will have smaller and smaller influence on the system evolution. Thus, in order to develop the quantum memory devices which work on the multi-atomic systems and to increase the quantum memory lifetime it can be important to optimize the number of particles and interaction between them.

The concept of thermalization offered in this article may help searching for a more complete suppression of decoherence in the multi-atomic quantum memory systems.

**Acknowledgments**

This work is supported by the Russian Foundation for Basic Research through the Grant No. 14-02-00903/14.